\documentstyle[11pt,aaspp4]{article}

\begin{document}

\title{The CTIO Prime Focus CCD: System Characteristics from 1982-1988}

\author{Nicholas B.\ Suntzeff\altaffilmark{1} and Alistair R. Walker}

\affil{Cerro Tololo Inter-American
Observatory\altaffilmark{2}, National Optical Astronomy Observatories\\
Electronic mail: nsuntzeff@noao.edu, awalker@noao.edu}

\altaffiltext{1}{On sabbatical leave at the Dominion Astrophysical 
Observatory, Herzberg Institute of Astrophysics, National Research Council,
5071 W. Saanich Road, Victoria, B.C. V8X 4M6, Canada}

\altaffiltext{2}{The National Optical Astronomy Observatories are operated
by the Association of Universities for Research in Astronomy, Inc., under
contract with the National Science Foundation.}

\centerline{RUNNING HEAD: The PFCCD System at CTIO}
\centerline{Send proofs to: N.B.\ Suntzeff, Cerro Tololo Inter-American
Observatory,  Casilla 603, La Serena, Chile}

\begin{abstract}

The CTIO Prime Focus CCD instrument with an RCA CCD was in operation at the
CTIO 4-m telescope for six years between 1982-1988. A large body of literature
has been published based on CCD images taken with this instrument. We review
the general properties of the now-retired PFCCD system to aid astronomers in
the interpretation of the photometric data in the literature.

\end{abstract}


\section{Introduction}

In October 1982 the KPNO prime focus CCD system (PFCCD) was transferred to
CTIO.  For six years, until de-commissioning in mid-1988, RCA \#1 CCD was the
only CCD available for direct imaging at the 4-m telescope.  The system is
described by Marcus et al. (1979), Goad (1980), and McGuire (1983), while from
a more astronomical perspective the faint galaxy study by Tyson and Seitzer
(1988) is recommended.

The detector is an RCA type SID52501 CCD, designed for use in high speed
(several Mpixel/second) 525-line TV applications.  The format is 320 x 512
pixels, each 1.2 mil ($30.48 \mu$) square with no dead bands.  It is a
thinned, back-illuminated device, with a silicon monoxide anti-reflection
coating.  The back surface is doped to prevent recombination near the surface,
which improves the ultraviolet and blue response.  The CCD is bonded to a
0.5mm-thick glass substrate, the front surface of which is anti-reflection
coated with magnesium fluoride.  The device data sheet and U.S patent
no. 4266334 (May 12 1981) can be consulted for further details. No definitive
description of the RCA CCD's has ever appeared in the literature.

This type of RCA CCD is known  as a ``first generation'' RCA  CCD, and has the
general characteristics of high readout noise (at least 70 electrons) and poor
vertical (parallel) charge transfer   at very low  light levels.    The output
amplifier emits light during readout.   Since RCA \#1  was used exclusively at
the CTIO   4-m prime  focus,   these defects  were   of little  importance for
broad-band (eg $BVRI$) imaging.  The  ``second (and final) generation'' of  RCA
CCD's employed at CTIO has 40 electrons readout noise, better low level charge
transfer, but a higher  radiation event rate  resulting  from a change  in the
glass substate material.

The CCD controller has been described in detail elsewhere (see above
references).  It incorporates a temperature controller set to --105C, which
results in a dark rate of approximately 70 electrons pixel$^{-1}$ hr$^{-1}$
when sampled well away from the glowing output amplifier.  The CCD output was
digitized to 15 bits, at a gain of 10.1 electrons per digital unit.  The data
were written as 16-bit unsigned integers. The 352 (320 CCD, 32 overscan) x 512
pixel raw images required 8 seconds for readout, with a noise of 85 electrons
pixel $^{-1}$ rms (root-mean-square).  During the normal on-site processing of
bias subtraction and flat-field division, it was normal to trim off several of
the bright left-hand columns.  The control computer was originally a PDP 11/23
with 256K memory, 77 MB Winchester disk, and a 1600 bpi 9-track magnetic tape
drive.

Apart from the low level charge transfer problem, many tests over the several
years of use at CTIO demonstrated that RCA \#1 was linear to $\sim 0.5\%$ or
better over the full range of the ADC, and saturation did not occur until
$\sim 400,000$ electrons.

The PFCCD system viewed a 3\arcmin\ x 5\arcmin\ field roughly 6.5\arcmin\ west
of the telescope optical axis, with the long axis oriented east-west.  The
PFCCD was used with the 4-m telescope doublet corrector, which is made of
fused silica with magnesium fluoride anti-reflection coatings.  This corrector
covers a field diameter of 17\arcmin.  Ray tracings show that at the
(off-axis) position of the CCD the corrector produces images with 0.4 arcsec
75\% enclosed energy. Qualitatively, these images have tight cores and
low-level wings, with little variation as a function of wavelength.  The image
scale with RCA CCD is $0.59 \pm 0.01$ arcsec/pixel, corresponding to $19.3 \pm
0.3$ arcsec/mm.

The first light of the instrument at CTIO was the night of 6 April 1982.  The
first scheduled run went to P. Seitzer and H. Butcher, which started on 12
April 1982.  The last scheduled run with the RCA \#1 CCD PFCCD was 25 March
1988.  Roughly 180 runs on 440 nights were scheduled with this instrument.
About 25\% of the observer time on the 4-m telescope (not used for engineering
purposes) went to the PFCCD during this period.  The most popular use of the
instrument was broad-band imaging in a subset of the $UBVRI$ Johnson and
Kron-Cousins photometric systems.  While the PFCCD was quite powerful for
broad-band imaging where sky-limited exposures were achieved in a few minutes,
the very high readout noise of the CCD made this system less attractive for
narrow-band imaging.

In this short contribution, we record transmission measurements for the most
popular $UBVRI$ filters used in by the PFCCD, together with quantum efficiency
measurements for RCA \#1. The latter measurement will be useful for those
astronomers trying to reconstruct photometric color terms for non-standard
filters. In addition, these tables in conjunction with spectrophotometric
atlases of standard stellar spectra (see for example, Massey et al. 1988,
Massey and Gronwall 1990, Hamuy et al. 1992, 1994) can be used to study the
effects of non-stellar flux distributions on the transformation of natural to
standard photometric systems. It is well-known that non-stellar flux
distributions can lead to systematic errors in the transformed photometry. For
instance, Suntzeff et al (1988) found systematic differences of up to 0.4
magnitudes in the $I$ magnitude of SN1987A that were explained by slight
differences in the $I$ filter convolved with the supernova flux distribution
(Bessell 1983, Menzies 1989).

\section{$UBV(RI)_C$ Filter Transmissions}
 
The original $BVRI$ filters used with the PFCCD system were a set of
interference filters obtained as part of a bulk order organized by J. Mould,
then at KPNO.  Two sets were allocated to the PFCCD, one as spare. The
interference $BV$ filters were rarely used. Instead, a set of glass filters
locally known as ``CTIO glass set \#2'' was used.  On 75\% of the PFCCD runs,
the \#2 $BV$ glass set was installed.

The CTIO glass set \#2 $BV$ filters are 2x2 inch square filters, 4mm thick,
and cemented with lens bond M62. The filter recipes are: $B$, BG12/1mm +
GG385/2mm + BG18/1mm; and $V$, GG495/2mm + BG18/2mm. The $B$ filter is
identical to the $B$ filter recommended (for photocathodes) by Bessell (1979),
and the $V$ is very similar to his $V$ filter. Some astronomers referred to
the CTIO glass set \#2 $BV$ filters as the ``Bessell'' set, although this is
only strictly true for the $B$ filter.  The interference $(RI)_C$ set was
generally known as the $R$34 and $I$34 filters or the ``Mould'' set. The
filters in this set are 2x2 inches and 3mm thick.

The 2x2 inch $U$ filters were less standard. The original $U$ filter was a 4mm
thick filter with the recipe: UG2/1mm + BG38/1mm + WG295/2mm.  Because it
suffered from red-leak, in 1984 an interference $U$ filter was purchased.
This was locally called the ``$U_{\rm CTIO}$-new'' filter or the
``3650\AA/600\AA'' filter, and was 6mm thick.  The filter only had 35\% peak
transmission, and in late 1985, this filter was replaced by a $U$ filter made
from a 1mm UG2 glass filter bonded to an optical cell containing copper
sulfate (80\%) solution, as recommended by Bessell (1976).  This final filter
had three times the throughput of the interference filter. The filter was
9.3mm thick.  Three of these filters were made by D. Hamilton for use at
CTIO. They are locally known as the ``$U$-Hamilton'' filters. Users of this or
any other liquid cell $U$ filter should note the warning in Bessell (1976)
that this type of cell can drastically change its transmissivity through very
small levels of ferric ion contamination.

The $UBV(RI)_C$ filters were measured in the Optics Laboratory on Cerro Tololo.
A double pass Oriel spectrometer feeds near monochromatic light at f/13.5
through the filter and then to either an S-5 or S-1 photomultiplier, depending
on the wavelength range to be measured.  At each wavelength both the filter
and the ``straight-through'' response are measured.

The transmission curves for the \#2 $BV$ glass set were measured in 1980 and
1989. These curves are plotted in Figure~\ref{x1} and the 1989 curves are
given in Table~\ref{y1} . The figure shows that there has been little change
in the transmissivity properties of this set over a 9-year period.

The transmission curves for the $RI$34 set were measured in 1983 and 1989. In
addition, we also have a curve for $R$34 from 1987. These curves are plotted
in Figure~\ref{x2} and the the 1989 curve is given in Table~\ref{y2}. Note in
this case there was a small shift in the transmission region of the
interference filters, especially in $I$34.

The $U$ transmission curves are plotted in Figure~\ref{x3}. In the left panel
we show the $U$-Hamilton \#1 filter as measured in 1986 and 1989. In the right
panel we show the ``$U_{\rm CTIO}$-new'' interference filter as measured in
1984 and 1989. Both filters remained quite stable during these time periods.
The 1989 curves are given in Table~\ref{y3}.

\subsection{Some comments on color terms}

Most users of the PFCCD system relied on the $UBV(RI)_C$ standard star lists
of Landolt (1973, 1983, 1992) and Graham (1982). There is a useful extension
of some of the Landolt fields in Stetson and Harris (1988).  These fields are
suited for CCD photometry since more than one standard can be fit on a frame.
Careful observers also chose stars from Menzies et al. (1989) as well as
Graham (1982) to include some stars that culminate between airmass 1.0 and
1.2, which are missing from the equatorial Landolt lists.

The bright limit for 1 second exposures was about $V=9.6$ for the PFCCD
system. Exposures this short must be corrected for the shutter timing
errors. The actual time the shutter was open was $t + \delta$ where $\delta$
was 10ms.

A large number of papers have been published based on the PFCCD photometry,
mostly in the $BV$ system. A number of these papers have presented the color
transformations from the natural to standard system (e.g. Aaronson et
al. 1984, Da Costa 1985, Smith et al. 1986, Hesser et al. 1987, McClure et
al. 1987, Bolte 1987). Most of the papers solve for a linear color
transformation in the ``traditional'' form of $M = M(m_0,c_0)$ where $M$ is
the transformed magnitude or color in the standard system, while $m_0$ and
$c_0$ are the observed magnitude and color on the natural system corrected for
extinction (Hardie 1962). For 11 different runs between 1982 and 1987, we find
the following mean linear color transformations:

\begin{displaymath}
V   = a + v_0 - 0.013(\pm 0.011)(b-v)_0 
\end{displaymath}
\begin{displaymath}
B-V = b + 1.092 (\pm 0.018) (b-v)_0 
\end{displaymath}

\noindent
where the error listed is the standard deviation (not mean error) of the 11
values.  It is our experience, and is verified by Stetson et al. (1989) that
the $BV$ transformations are non-linear if very red stars are included. We
recommend the transformation scheme suggested by Harris et al. (1981) and
Stetson and Harris (1988) where the non-linear transformation is written as $m
= m(M,C,X,...)$ where $M$ and $C$ are the standard colors, $X$ is the airmass,
and $m$ is the observed magnitude. These techniques also allow the observer to
solve for color terms on nights of thin cirrus, or include data where only a
single filter was employed.
 
\section{Quantum Efficiency of RCA \#1}
 
The quantum efficiency of CTIO CCD RCA \#1 was measured with respect to a
calibrated photo-diode using an apparatus (R2D2) designed at CTIO by Dr B.
Atwood.  In this apparatus, an Oriel spectrophotometer feeds a monochromatic
beam along a fiber-optic cable, into an integrating sphere which, via relay
lenses and a beam-splitter, focuses the light onto both the calibrated diode
and the CCD.  Measurements of relative sensitivity of the CCD and diode are
made as a function of wavelength, usually at 100\AA\ intervals.  The
resolution, governed by the slit width, is normally set to 50\AA. The CCD and
photo-diode are then interchanged with respect to the incoming beam and a
second series of measurements taken.  The two sets of measurements can then be
combined in order to remove instrumental effects.  A deuterium lamp,
ultraviolet fiber (fused silica) and ultraviolet - integrating sphere are
normally used for the 3000 - 4500\AA\ wavelength region.  A quartz lamp is
used for the region 4000 - 10000\AA, with its own integrating sphere which is
fed by a trifurcated fiber cable in order to improve the evenness of
illumination.  All measurements are made in the first order of the grating,
with a long pass filter inserted to block the second order for measurements at
wavelengths longer than 6000\AA.  Control of the spectrometer and recording of
data proceeds automatically under computer control.

Repeated measurements give us confidence that the {\it relative} quantum
efficiencies measured using R2D2 are accurate to better than 2\%.  However we
prefer to be conservative with respect to the overall {\it absolute} scale,
and will assign an error of 10\%, even though R2D2 is designed to eliminate
sources of systematic error as much as possible.  We find that the absolute
peak quantum efficiency of RCA \#1 is somewhere in the range 57 - 69\%.  The
quantum efficiency of RCA \#1 is plotted in Figure~\ref{x4} and listed in
Table~\ref{y4}.

The shape of the spectral response for RCA CCD's depended critically on the
amount of thinning.  Generally, ``second generation'' devices were not thinned
as much as the ``first generation'' CCD's and consequently have poorer
response in the ultraviolet but better red response.  However the degree of
thinning during manufacture was evidently not well controlled and some ``first
generation'' devices also had poor ultraviolet response.  RCA \#1 has just the
opposite; excellent ultraviolet but poor red response.  It was also renowned
for the strength of the fringes seen when illuminated monochromatically, such
as from night sky emission lines when imaging in the V, R and I bands.  These
fringes are stronger the thinner the CCD, at least for the RCA CCD's.

\section{Other comments}

While the filter transmission and CCD quantum efficiency dominate the final
throughput curve for the system, there are some other effects that will modify
the final number of detected photons per spectral element. Besides the obvious
case of interstellar extinction (see Cardelli et al. 1989 for a particularly
convenient form of this law) and atmospheric extinction (an average law for
CTIO is given by Gutierrez-Moreno et al. 1986), light will be attenuated by
the aluminum reflection from the primary mirror and the passage through the
doublet corrector. The reflectivity of aluminum is conveniently summarized by
Smith, et al (1985) but it should be noted that this curve is for an
idealized, fresh aluminum surface.  A plot of the mirror reflectivity for the
freshly aluminized Canada-France-Hawaii Telescope mirror is given by Magrath
(1994).  The reflectivity of fresh aluminum is very high and uniform
throughout the optical, except for an interband absorption feature at about
8500\AA\ which is about 1500\AA\ wide with a maximum absorption of about 10\%.

We have laboratory information on the transmissivity of the prime focus
corrector, which is a fused silica, air spaced doublet system. The glass
transmissivity of the elements, however, is expected to be flat throughout the
optical.

A common point of confusion for observers with the old PFCCD system was the
coordinate center for the CCD image. In almost all cases, the coordinates
written in the header of the image referred to position of the telescope
optical axis, and {\it not} the CCD position. The true CCD position was $\sim
6.5\arcmin$ west of the telescope optical axis - that is, one must subtract
6.5\arcmin\ (in units of RA) from the headers stored in the images to recover
the CCD center. In a few cases, the observers re-zeroed the telescope pointing
to the CCD center, but this was not recommended because the telescope pointing
model would be less accurate.

\acknowledgments

We wish to thank Dr. Ivan King for urging the publication of the basic
characteristics of the CTIO PFCCD system.  We would like to thank R. Gonzalez
and G. Martin for measuring the filter responses. We thank P. Stetson, G. Da
Costa, M. Bolte, and M. Dickinson for communicating their experience with the
PFCCD system.  G. Jacoby provided the ray tracings for the doublet corrector.
We thank B. Magrath for providing information on aluminum reflectivities.

\clearpage

\section*{References}
\begin{verse}
Aaronson, M., Schommer, R.A., and Olszewski, E. 1984, \apj, 276, 221\\
Bessell, M.S. 1976, \pasp, 88, 557\\
Bessell, M.S. 1979, \pasp, 91, 589\\
Bessell, M.S. 1983, \pasp, 93, 507\\
Bolte, M. 1987, \apj, 315, 469\\
Cardelli, J. A., Clayton, G. C., and Mathis, J. S. 1989, \apj, 345, 245\\
Da Costa, G.S. 1985, \apj, 291, 230\\
Goad, L. 1980, Proc. SPIE, 264, 136\\
Graham, J.A. 1982, \pasp, 94,244 \\
Gutierrez-Moreno, A., Moreno, H., and Cortes, G.   1986, \pasp, 98, 1208\\
Hamuy, M., Walker, A.R., Suntzeff, N.B., Gigoux, P., Heathcote, S.R.,
 and Phillips, M.M. 1992, \pasp, 104, 533\\
Hamuy, M., Suntzeff, N.B., Heathcote, S.R., Walker, A.R., Gigoux, P.,
 and Phillips, M.M. 1994, \pasp, 106, 566\\
Hardie, R.H. 1962, in Astronomical Techniques, ed. W.A. Hiltner, 
 (Chicago: University of Chicago Press), p. 178\\
Harris, W. E., Fitzgerald, M. P., and Reed, B. C. 1981, \pasp, 93, 507\\
Hesser, J.E., Harris, W.E., VandenBerg, D.A., Allwright, J.W.B., Shott, P.,
 and Stetson, P.B. 1987, \pasp, 99, 739\\
Landolt, A.U. 1973, \aj, 78, 959\\
Landolt, A.U. 1983, \aj, 88, 439\\
Landolt, A.U. 1992, \aj, 104, 340\\
Massey, P., Strobel, K., Barnes, J.V., and Anderson, E. 
 1988, \apj, 328, 315\\
Massey, P. and Gronwall, C. 1990, \apj 358, 344\\
Marcus, S.L., Nelson, R.E., and Lynds, C.R. 1979, Proc. SPIE, 172, 207\\
McClure, R.D., VandenBerg, D.A., Bell, R.A., Hesser, J.E., and Stetson, P.B.
 1987, \aj, 9, 1144\\
 McGuire, T.E. 1983, \pasp, 95, 919\\
Magrath, B. 1994, CFHT Info. Bull, 30, 15\\
Menzies, J. W., Cousins, A. W. J., Banfield, R. M., and Laing, J. D.
 1989, South Af. Astron. Obs. Circ., 13, 1\\
Menzies, J. W. 1989, \mnras, 237, 21 \\
Smith, G.H., McClure, R.D., Stetson, P.B., Hesser, J.E., and Bell, R.A. 
 1986, \aj, 842 \\
Smith, D.Y., Shiles, E., and Inokuti, M. 1985, in Handbook of 
 Optical Constants of Solids, ed. E.D. Palik (Orlando: Academic Press), 
 p. 369\\
Stetson, P.B., and Harris, W.E. 1988,\aj, 96, 909 \\ 
Stetson, P.B., VandenBerg, D.A., Bolte, M., Hesser, J.E., and Smith, G.H. 
 1989, \aj, 97, 1360\\
Suntzeff, N.B., Hamuy, M., Martin, G., G\'omez, A., and Gonz\'alez, R. 1988, 
 \aj, 96, 1864\\
Tyson, J.A. and Seitzer, P. 1988, \aj, 335, 552\\
\end{verse}

\clearpage

\figcaption[]{Transmission curves for the CTIO glass set \#2
$BV$ filters. The dashed line represents the filter transmission for 1980, and
the solid line for 1989.  Note the constancy of the curves, as would be
expected for glass filters. \label{x1}}

\figcaption[]{Transmission curves for the $RI$34 interference filter set. 
The dashed line represents the filter transmission for 1983, and the solid
line for 1989.  An additional curve from 1987 is plotted for $R$34 as a dotted
line.  Note the small changes in the curves, especially for $I$34.
\label{x2}}

\figcaption[]{Transmission curves for the $U$ filters. 
The left panel shows the curves for the ``$U$-Hamilton'' filters, which are
the most used $U$ filters on the PFCCD system. The dashed curve represents the
filter transmission in 1986, and the solid curve in 1989. The right panel
shows the curves for the ``$U_{\rm CTIO}$-new'' interference filter.  The
dashed curve represents the filter transmission in 1984, and the solid curve
in 1989.
\label{x3}}

\figcaption[]{The quantum efficiency curve for the RCA CCD \#1, which was used
in the PFCCD system from 1982-88. \label{x4}}

\clearpage
 
\begin{table}
\dummytable\label{y1}
\end{table}

\begin{table}
\dummytable\label{y2}
\end{table}

\begin{table}
\dummytable\label{y3}
\end{table}

\begin{table}
\dummytable\label{y4}
\end{table}

\noindent{\sc TABLE}~\ref{y1}. Transmission Curves for
 \#2 $BV$ CTIO Glass Filter Set

\noindent{\sc TABLE}~\ref{y2}. Transmission Curves for
 $RI$34 CTIO Interference Filter Set

\noindent{\sc TABLE}~\ref{y3}. Transmission Curves for 
 CTIO $U$ Filters

\noindent{\sc TABLE}~\ref{y4}. Quantum Efficiency Curve for RCA \#1 CCD

\begin{figure}
\plotone{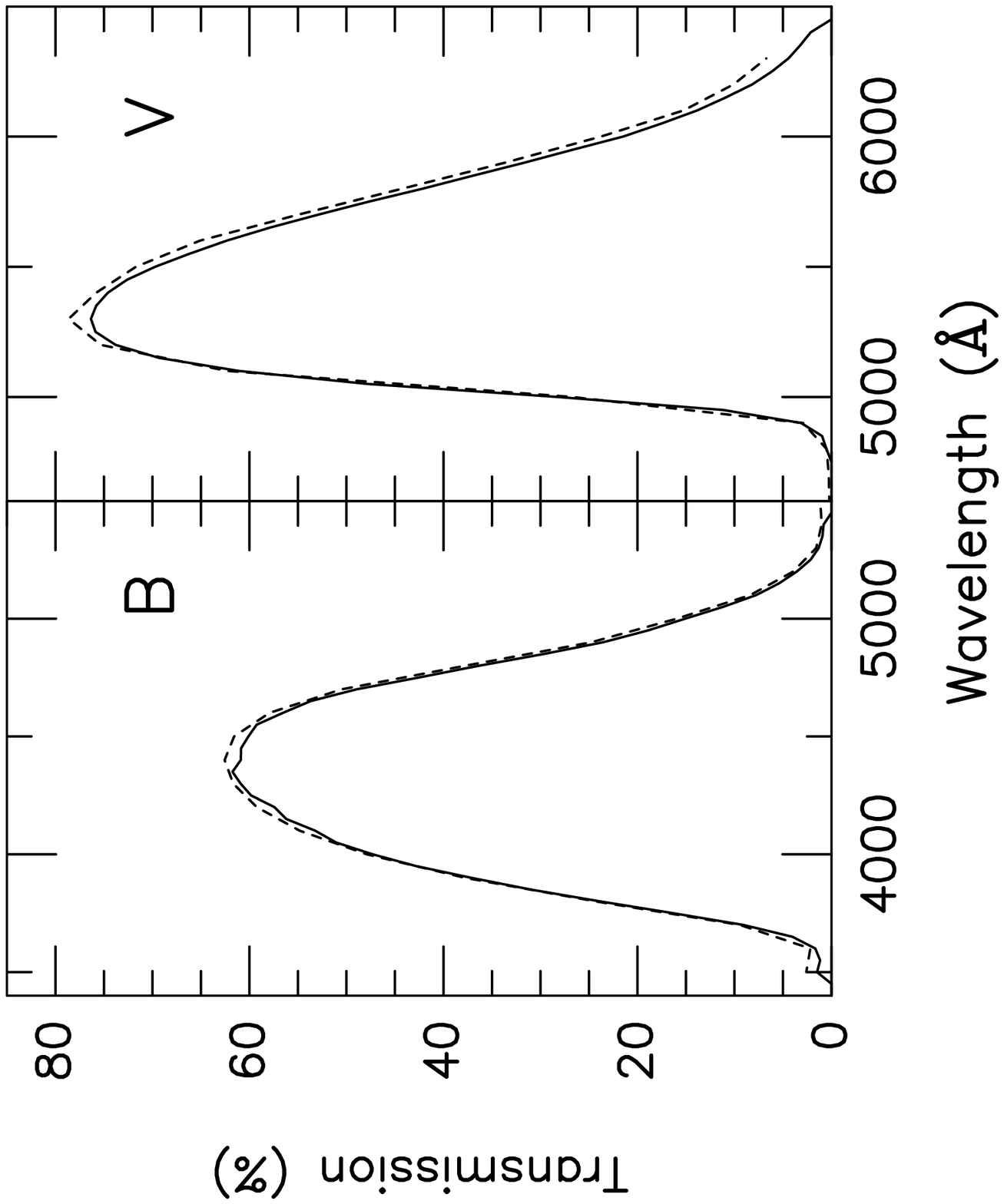}
{\center Suntzeff and Walker. Figure~\ref{x1}}
\end{figure}

\begin{figure}
\plotone{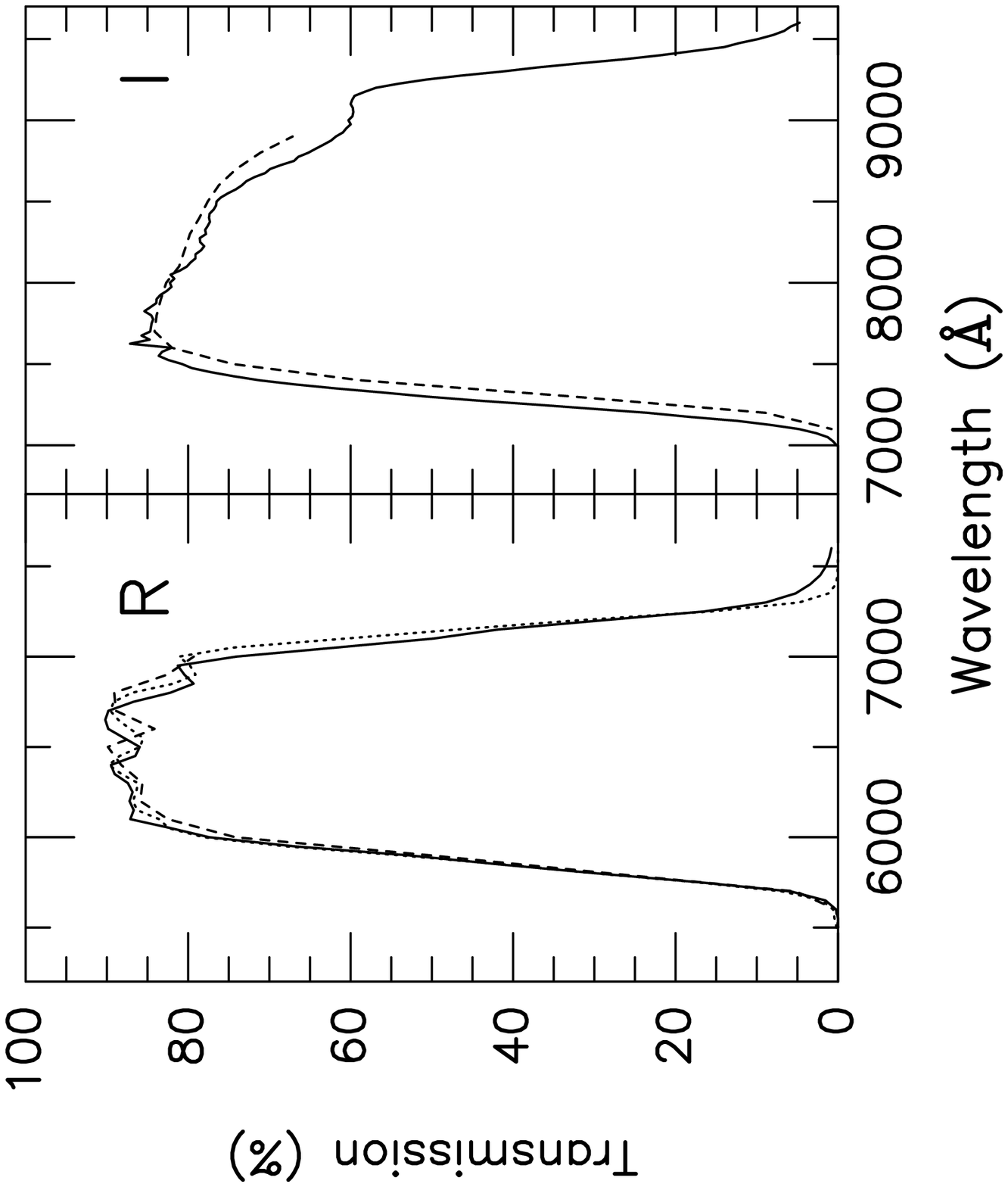}
{\center Suntzeff and Walker. Figure~\ref{x2}}
\end{figure}

\begin{figure}
\plotone{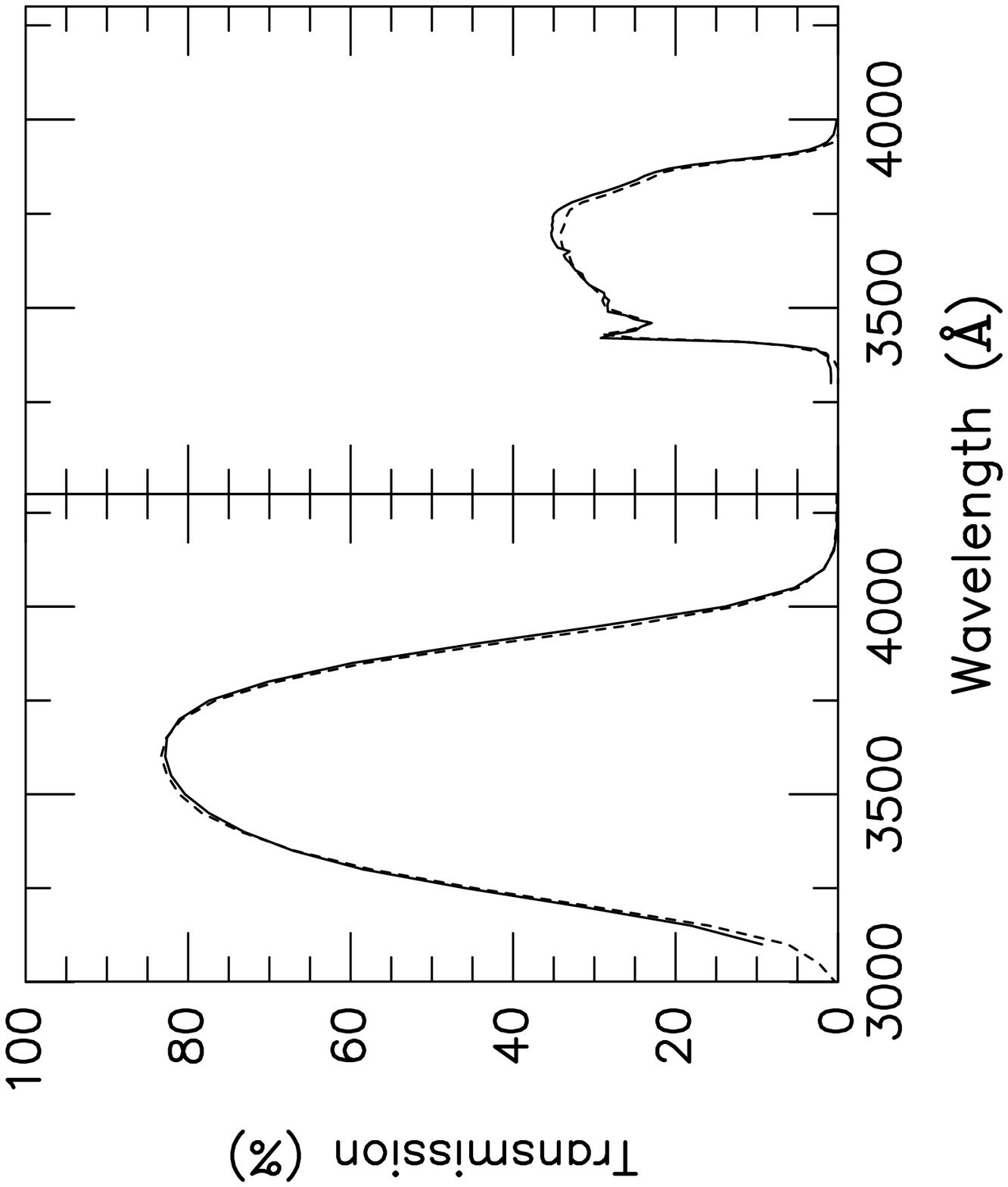}
{\center Suntzeff and Walker. Figure~\ref{x3}}
\end{figure}

\begin{figure}
\plotone{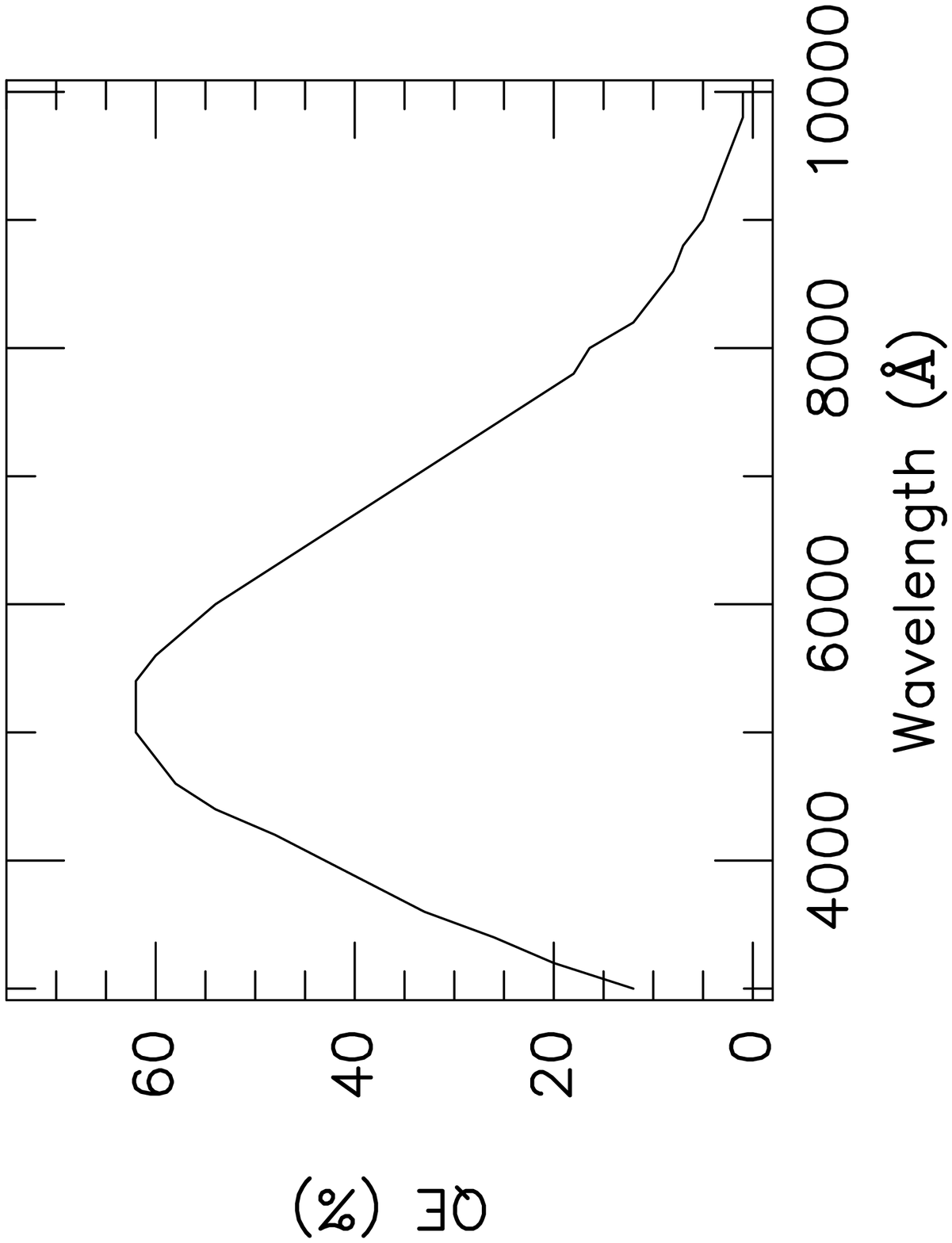}
{\center Suntzeff and Walker. Figure~\ref{x4}}
\end{figure}

\end{document}